\documentclass[twocolumn]{aastex63}

\usepackage{amsmath}
\usepackage{color,soul}
\usepackage{natbib}
\usepackage{float}

\newcommand{\num}{65 } 
\newcommand{\zhi}{\ensuremath{z\sim0.7}}
\newcommand{\zlo}{\ensuremath{z\sim0}}
\newcommand{\ang}{\ensuremath{\mathring{\rm A}}}
\usepackage{lineno}

\accepted{July 9, 2021}
\shorttitle{Stellar MZR at \texorpdfstring{$z\sim0.7$}{z~0.7}}
\shortauthors{Aliza G. Beverage}

\begin{document}





\title{Elemental Abundances and Ages of $z\sim0.7$ Quiescent Galaxies on the Mass-Size Plane: \\ Implication for Chemical Enrichment and Star-Formation Quenching}

\correspondingauthor{Aliza Beverage}
\email{abeverage@berkeley.edu}

\author[0000-0002-9861-4515]{Aliza G. Beverage}
\affiliation{Department of Astronomy, University of California, Berkeley, CA 94720, USA}
\author[0000-0002-7613-9872]{Mariska Kriek}
\affiliation{Department of Astronomy, University of California, Berkeley, CA 94720, USA}
\author[0000-0002-1590-8551]{Charlie Conroy}
\affiliation{Center for Astrophysics \textbar\ Harvard \& Smithsonian, Cambridge, MA,
02138, USA}
\author[0000-0001-5063-8254]{Rachel Bezanson}
\affiliation{Department of Physics and Astronomy, University of Pittsburgh, Pittsburgh, PA 15260, USA}
\author[0000-0002-8871-3026]{Marijn Franx}
\affiliation{Leiden Observatory, Leiden University, P.O.Box 9513, NL-2300 AA Leiden, The Netherlands}
\author[0000-0002-5027-0135]{Arjen van der Wel}
\affiliation{Sterrenkundig Observatorium, Universiteit Gent, Krijgslaan 281 S9, B-9000 Gent, Belgium}




\begin{abstract}
We present elemental abundances and stellar population ages for \num massive quiescent galaxies at $0.59\leq z\leq0.75$ from the LEGA-C survey. The abundance patterns and ages, derived from full-spectrum modeling, are examined as a function of stellar mass ($M_*$) and size (i.e., half-light radius; $R_e$). We find that both [Mg/H] and [Fe/H] do not vary with stellar mass but are correlated with $M_*/R_e$ for quiescent galaxies with $M_*>10^{10.5}$\;M$_\odot$. Thus, at fixed mass, compact quiescent galaxies are on average more metal rich. This result reinforces the picture that supernova feedback and gravitational potential regulate chemical enrichment. [Mg/Fe] does not vary with $M_*$ or $M_*/R_e$, but there is a marginal positive relation between age and mass. Our results support low-redshift findings that more massive galaxies form their stars at earlier times. However, in contrast to low-redshift studies, star formation timescale does not appear to depend on mass or size. We also compare the mass-[Fe/H] and mass-[Mg/H] relations to stacks of quiescent galaxies at $z\sim0$ and find that both relations increase by $\sim0.2$\;dex over the past 7 Gyr. Furthermore, at \zhi\ we find a clear trend with age, such that older quiescent galaxies have lower metallicities. Both results can be explained by a chemical evolution model in which galaxies quench via gas removal. Future work, in particular with JWST/NIRSpec, will extend this analysis to higher redshifts, allowing us to fully exploit abundance patterns to study the formation histories of quiescent galaxies.
\end{abstract}


\section{Introduction}
\label{sec:intro}

The metallicity of a galaxy is a fundamental property sensitive to many complex evolutionary processes such as metal production and enrichment, removal of enriched gas via galactic winds, and the accretion of circum/intergalactic gas. It is well established that metallicity is correlated with galaxy mass, wherein galaxies with larger stellar mass are more metal rich \cite[e.g.,][]{lequeux_reprint_1979,tremonti_origin_2004}. This so-called mass-metallicity relation (MZR) has been confirmed to extend over five decades in stellar mass, flattening at the highest masses \citep[e.g.,][]{tremonti_origin_2004,gallazzi_ages_2005,kirby_universal_2013}. The MZR holds for both the gas-phase metallicity, as well as the stellar metallicity, and has been observed in both star-forming and quiescent populations. The origin of this relation is still debated, but has historically been attributed to the strength of a galaxy's potential well; galaxies with larger stellar masses, and thus larger escape velocities, are better at retaining metal-enriched gas \citep[e.g.,][]{larson_effects_1974,dekel_origin_1986,tremonti_origin_2004}. 

Interestingly, studies of low-redshift galaxies have found that metallicity has a secondary dependence on galaxy size, such that at a fixed mass, smaller galaxies have higher metallicities. This size dependence, which has been found in both star-forming \citep[e.g.,][]{ellison_clues_2008,scott_sami_2017,li_sdss-iv_2018,deugenio_gas-phase_2018} and quiescent \citep{mcdermid_atlas3d_2015,barone_sami_2018,li_sdss-iv_2018} populations, reinforces the importance of the gravitational potential in regulating metallicity; at a given mass, smaller galaxies have steeper potential wells.

\begin{figure*}
    \centering
    \includegraphics[width=\textwidth]{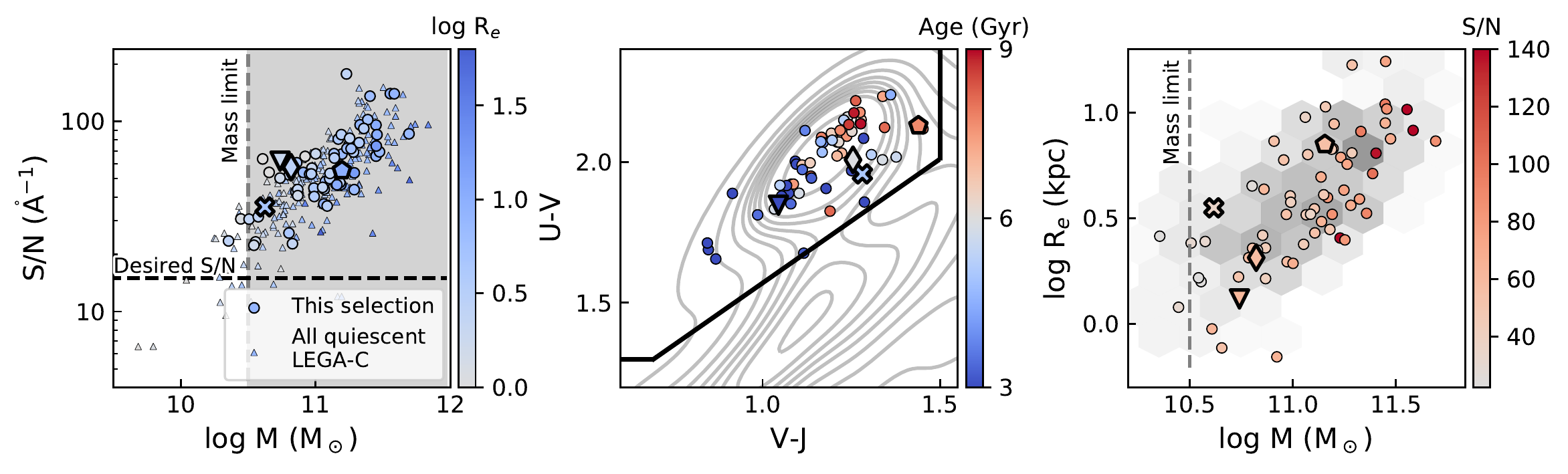}
    \caption{\textit{Left:} the distribution of stellar mass and S/N at rest-frame 5000\,$\mathring{\rm A}$ of all available quiescent LEGA-C galaxies in the redshift range $0.59<z<0.75$ colored by their half-light radius. Galaxies included in the final selection are identified. The 95\% mass-limit ($\log M_*/M_\odot = 10.5$) and required rest-frame S/N\,($=15\,\mathring{\rm A}^{-1}$) are marked. Not all galaxies with S/N\,$>15$ are included in the final sample, as we require that the spectra cover $\lambda = 4800-5430\,\mathring{\rm A}$. \textit{Middle:} the selected sample in UVJ space colored by stellar age, as derived from \texttt{alf}. The contours represent the full LEGA-C sample. \textit{Right:} the selected sample in mass-size space colored by S/N at rest-frame 5000\,$\mathring{\rm A}$. The density map underneath shows the quiescent LEGA-C sample from the left panel. In all panels we highlight four galaxies using different symbols that correspond to the galaxies in Fig.\;\ref{fig:spec}.}
    \label{fig:sn}
\end{figure*}

Whereas the star-forming MZR has been extensively studied out to $z\sim3.5$ \citep[e.g.,][]{erb_mass-metallicity_2006,onodera_ism_2016,sanders_mosdef_2020}, the quiescent MZR has only been measured out to $z\sim0.7$ \citep{gallazzi_charting_2014}. Measuring robust metallicities of distant quiescent galaxies is extremely difficult and requires ultra-deep continuum spectroscopy, as it relies on faint absorption lines shifted to near-IR (NIR) wavelengths. Consequently, the few existing studies beyond low-redshift have large uncertainties and/or are based on stacked spectra \citep{gallazzi_charting_2014,choi_assembly_2014,leethochawalit_evolution_2019}. 

In this Letter, we utilize the public Large Early Galaxy Astrophysics Census \citep[LEGA-C;][]{van_der_wel_vlt_2016,straatman_large_2018}, which has obtained ultra-deep continuum spectra ($\sim20$ hrs) for 1988 galaxies at $0.6<z<1$ in the COSMOS field using VIMOS on the Very Large Telescope with an average signal-to-noise ratio (S/N) of $\rm{S/N}\sim20\,\ang^{-1}$. We analyze the spectra using a full-spectrum modeling code. We then assess how the metal content of quiescent galaxies at $z=0.59-0.75$ relate to their stellar masses and sizes, and study how these relations evolve between \zhi\ and \zlo. Throughout this work we assume a flat $\Lambda$CDM cosmology with $\Omega_{\rm m}= 0.29$ and $H_{\rm 0} = 69.3$\;km\,s$^{-1}$\,Mpc$^{-1}$.

\begin{figure*}
    \centering
    \includegraphics[width=\textwidth]{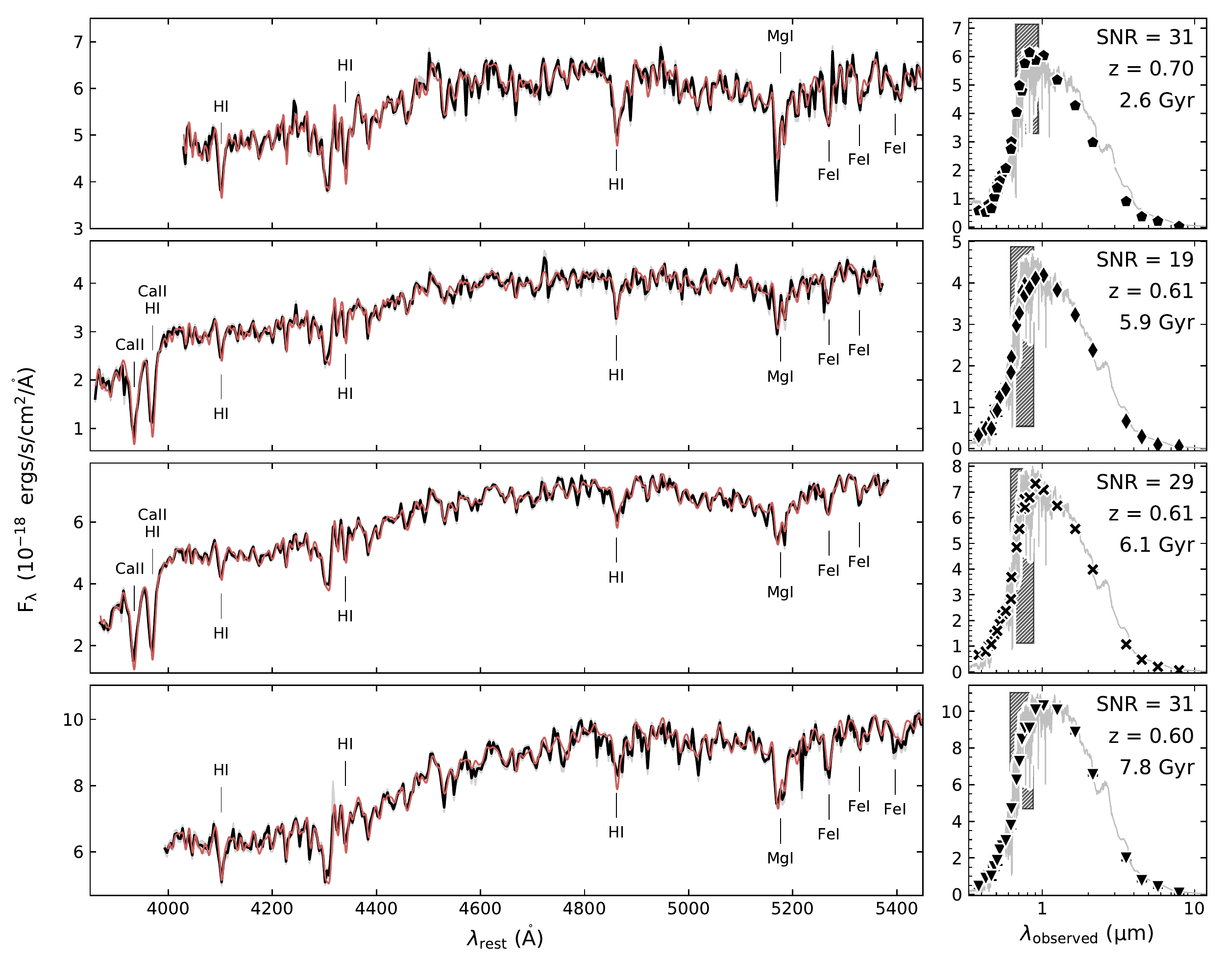}
    \caption{\textit{Left:} LEGA-C spectra for four representative galaxies (black) with 1$\sigma$ uncertainties in gray. For displaying purposes the spectra are binned by 10 pixels, corresponding to $\sim$3.5 rest-frame\;$\mathring{\rm{A}}$ per bin. The best-fit \texttt{alf} model is shown in red. \textit{Right}: the corresponding photometry along with the best-fit FSPS model. The gray boxes represents the extent of the left panels. The S/Ns of the spectra are reported at rest-frame 5000 $\mathring{\rm A}$ per rest-frame $\mathring{\rm A}$. The symbols used for the photometry match those in Fig.\;\ref{fig:sn}.}
    \label{fig:spec}
\end{figure*}

\section{Sample and Methods}
\label{sec:methods}

We use deep continuum spectroscopy from the public Large Early Galaxy Astrophysics Census (LEGA-C), a survey of 3600 galaxies at $0.6<z<1$ selected from the UltraVISTA K-band catalog by \cite{muzzin_public_2013}. The deep, high-resolution spectra were collected using the VIMOS multiobject spectograph at the ESO Very Large Telescope and have an average S/N\,$\sim20\,$\ang$^{-1}$. For more information on the survey design see \cite{van_der_wel_vlt_2016}. For details on the observation and data reduction see \cite{straatman_large_2018}. 

We select quiescent galaxies from the second LEGA-C data release (1922 targets), based on the rest-frame $U-V$ and $V-J$ colors, as prescribed in \cite{muzzin_public_2013}. From the 751 quiescent galaxies, we make a selection based on the quality and wavelength range of the spectra. First, to ensure reliable abundance pattern measurements, we require each spectrum to contain at least two FeI features and one Balmer line (H$\beta$), translating to a required rest-frame wavelength coverage of $4800\,\mathring{\rm A} < \lambda < 5430\,\mathring{\rm A}$. This wavelength selection effectively sets an upper limit on the redshift ($z=0.75$). We set the minimum redshift to $z=0.59$, thereby removing spurious $z\sim0.3$ galaxies. Second, we select galaxies with rest-frame S/N\;$> 15\,\mathring{\rm A}^{-1}$ as measured at $5000\,\mathring{\rm A}$. Finally, we visually inspect the 87 remaining spectra and remove five galaxies with obvious emission lines (which are likely active galactic nuclei (AGN) or chance alignments with a star-forming galaxy). The initial selection includes 82 quiescent galaxies for which we derive stellar masses and sizes.

\subsection{Galaxy Sizes and Stellar Masses}

We derive galaxy sizes from the \textit{HST} ACS COSMOS F814W images \citep{scoville_cosmic_2007} by fitting a single-component S\'ersic model with \texttt{GALFIT} \citep{peng_detailed_2010}. For each galaxy we make 25$''$ cutouts and mask out nearby sources. Similar to the procedure in \cite{van_der_wel_3d-hstcandels_2014}, the S\'ersic index is constrained to the values $0.7\leq n\leq8$. We define galaxy size ($R_e$) as the semi-major axis of the ellipse containing half of the total flux of the best-fit model.

We visually inspect the resulting best-fit S\'ersic models and residuals and remove one galaxy in a close pair and one with a disturbed morphology (likely due to a recent merger). Additionally, we remove three galaxies with \texttt{GALFIT}-derived S\'ersic indices that hit the $n=8$ constraint.

Next, we derive stellar masses by fitting UltraVISTA photometry with Flexible Stellar Population Synthesis \citep[\texttt{FSPS}][]{conroy_propagation_2009} templates using the \texttt{FAST} fitting code \citep{kriek_ultra-deep_2009}. We assume a delayed exponentially declining star formation history, the \cite{chabrier_galactic_2003} initial mass function (IMF), and the \cite{kriek_dust_2013} dust attenuation law.
We also assume solar metallicity to facilitate comparison to other studies. To test this last assumption, we take six metal-poor galaxies and refit their masses by fixing the template metallicities to the values derived from their spectra. The resulting masses do not change by more than 0.1\;dex. Next, we correct the stellar masses such that they are consistent with the \texttt{GALFIT} profiles by multiplying them by the ratio of the \texttt{GALFIT}-derived F814W flux and the interpolated F814W flux from the photometric catalog. On average, this procedure
increases the stellar mass by 4\%. Furthermore, we increase the masses by 0.1\,dex to correct for a systematic offset between the space-based 3D-\textit{HST} and ground-based UltraVISTA photometry \citep[see][]{mowla_cosmos-dash_2019}.



We determine a lower mass-limit using an iterative method. We first derive the relation between S/N and log\,$M/M_\odot$ by fitting to all available quiescent galaxies at $0.59<z<0.75$ (272; see the left panel of Fig.\;\ref{fig:sn}). Using this fit, we take all galaxies above a given cutoff mass and calculate what their S/N would be if they had a mass equal to the cutoff mass. We then determine the S/N below which 5\% of these scaled-down galaxies would not be included in the sample. We repeat this method, starting with a high cutoff mass and iterating down to lower masses. We find that at $M=10^{10.5}\,M_{\odot}$ the sample is 95\% complete for a S/N criterion of $15\,\mathring{\rm A}^{-1}$. We note that at this redshift LEGA-C is representative of the full galaxy population down to log\,$M/M_\odot=10.4$.

\begin{figure*}[!tp]
    \centering
    \includegraphics[height=0.8\textheight]{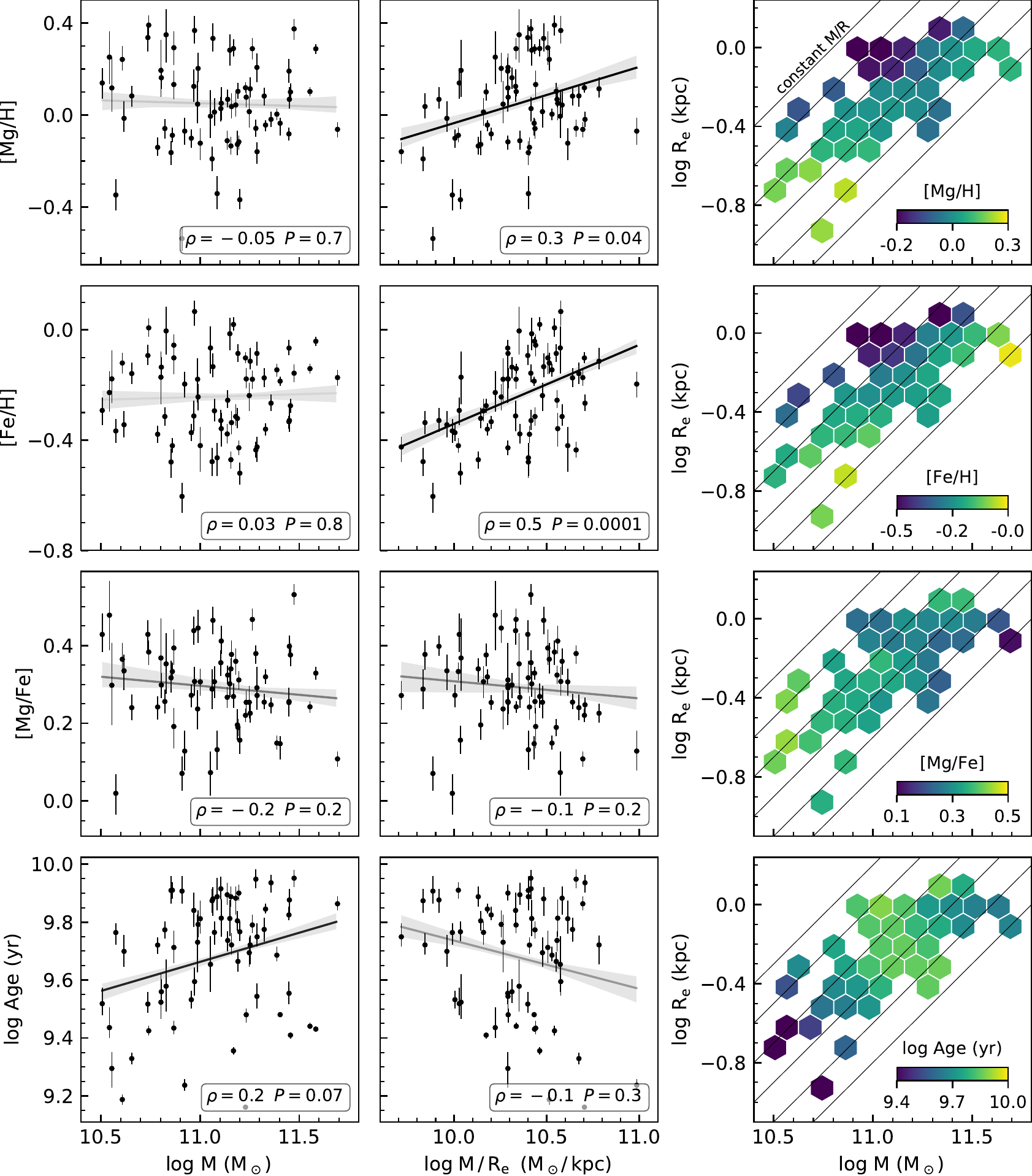}
    \caption{Correlation between [Mg/H], [Fe/H], [Mg/Fe], and stellar age with stellar mass (left) and $M_*/R_e$ (middle). The right column shows the parameters on the mass-size plane, averaged using the LOESS algorithm. In the bottom right corner of these panels we show the $P$-value from a Spearman correlation test. Correlations with $P$-values less than $P<0.05$ are considered significant. The best-fit linear relations are shown in each panel, with the shaded regions showing the $1\sigma$ confidence intervals. The transparency of the best-fit lines are a reflection of the \textit{P}-value. [Mg/H] and [Fe/H] show no trend with $M_*$ but are significantly correlated with $M_*/R_e$.}
    \label{fig:ab}
\end{figure*}

\subsection{Elemental Abundances}
\label{sec:alf}

We derive the elemental abundances for the remaining 77 galaxies using the full-spectrum absorption line fitter (\texttt{alf}) code \citep{conroy_stellar_2012,conroy_metal-rich_2018}. This method is preferred over the use of specific features (i.e., Lick indices) because it can leverage the entire spectrum and is better at dealing with (non-Gaussian) noise due to OH lines or instrumental artifacts. \texttt{alf} fits optical--NIR spectra of quiescent populations older than $\gtrsim1$ Gyr and uses libraries of isochrones and empirical stellar spectra \citep{sanchez-blazquez_medium-resolution_2006,choi_mesa_2016,villaume_extended_2017}, along with synthetic spectra covering a wide range of elemental abundances. In the analysis presented here, we use \texttt{alf} in \textit{simple} mode, which fits for 13 parameters; a single stellar age, velocity dispersion, velocity offset, metallicity scaling, and the abundances of 9 elements (Fe, O, C, N, Na, Mg, Si, Ca, and Ti). We assume the \cite{kroupa_variation_2001} IMF, fit using a single burst star formation history, and use metallicity-independent response functions. We also set a maximum age of 10 Gyr, which is the age of the universe at $z=0.75$ +1 Gyr. 



Given the flexibility of \texttt{alf}, we run many tests to ensure the robustness of our measurements. These tests include: fitting without the age-sensitive CaII HK lines, allowing for two bursts of star formation, and with and without emission. The results of these tests are all within error of the analysis presented here.

After fitting, we visually inspect the best-fit \texttt{alf} models and the corresponding corner plots. Galaxies with unconstrained ages or metallicities, or those that run up against the upper-limit age prior are removed from the sample. Many of these galaxies have obvious skyline contamination near key absorption features. After removing these ten galaxies, along with the two galaxies that fall below the mass limit (see the left and right panels of Fig.\;\ref{fig:sn}), we are left with the final sample of \num galaxies. 
The mean redshift of the sample is $z=0.66$. Fig.\;\ref{fig:sn} shows the distribution of the final sample in mass-S/N (left), UVJ (middle), and mass-size (right) space. Fig.\;\ref{fig:spec} shows LEGA-C spectra along with their corresponding best-fit \texttt{alf} models for four representative galaxies.

\section{Chemical abundances of \texorpdfstring{$\lowercase{z}\sim0.7$}{z0.7} quiescent galaxies}

\label{sec:size}
In this section, we investigate the dependence of elemental abundances and stellar population ages on galaxy mass and size for quiescent galaxies at \zhi. In the left and middle columns of Fig.\;\ref{fig:ab} we show [Mg/H], [Fe/H], [Mg/Fe], and age versus stellar mass and mass-to-radius ratio ($M_*/R_e$). 
We use $M_*/R_e$ instead of velocity dispersion $\sigma$ ($\propto M/R$), as the quiescent LEGA-C galaxies have been found to be partially rotationally supported \citep{bezanson_spatially_2018}. For each panel, we fit to all galaxies with a linear regression and measure their Spearman rank coefficient, where the confidence intervals are determined by perturbing each data point in the y-direction according to their uncertainties. Including the \textit{x}-direction uncertainties ($\sim$0.1\;dex) in the fits does not significantly change the results \citep[see][for motivation of mass uncertainties]{kriek_massive_2016}. In the right column we show the mean trends of the stellar population parameters on the mass-size plane, where the properties have been smoothed using the Locally Weighted Regression \citep[LOESS;][]{cleveland_locally_1988} python package \citep{cappellari_atlas3d_2013}.

We first consider [Mg/H] and [Fe/H]. Mg, an $\alpha$-element, is almost exclusively produced in the cores of massive stars and released via core-collapse supernovae. Given the short lifetime of massive stars, Mg is almost instantaneously released and recycled and is thus a good tracer of the overall enrichment. Fe-peak elements are also produced in the cores of massive stars, but are predominantly forged during Type Ia supernovae. The latter descend from remnants of long-lived, low-mass stars, and as such, Type Ia supernovae products are only recycled if star formation is still ongoing. Thus, [Fe/H] depends both on the overall metal enrichment \textit{and} the duration of star formation. Interestingly, as seen in the first two rows of Fig.\;\ref{fig:ab}, both [Fe/H] and [Mg/H] correlate with $M_*/R_e$, but not with $M_*$, despite their different enrichment mechanisms. These trends are especially apparent in the mass-size panels.

Our results are consistent with studies at low-z that find a flattening of the quiescent MZR at the highest masses \citep{gallazzi_ages_2005,peng_strangulation_2015}, but show a strong increasing trend with $M_*/R_e$ \citep{barone_sami_2018}. Studies of high-mass quiescent galaxies at $z\sim0.5$ find similarly shallow mass-[Fe/H] and mass-[Mg/H] relations for stacks of galaxies \citep{choi_assembly_2014} and cluster galaxies \citep{leethochawalit_evolution_2019}. \cite{gallazzi_charting_2014} also find a slope for the MZR consistent with zero at $z\sim0.7$ using Lick indices; this is the only study of the quiescent MZR at a comparable redshift to this study.

Our results show, for the first time, that size plays an important role in regulating the elemental abundances of massive quiescent galaxies beyond $z\sim0.1$; while the trends with $M_*$ are mostly flat, we find strong correlations with $M_*/R_e$. This result implies that chemical enrichment is regulated by the strength of the potential well, as $M_*/R_e$ is a better tracer of the gravitational potential than just $M_*$ ($\Phi\propto M/R$). The flattening of the abundances with $M_*$ is thought to be due to minor mergers, as more massive galaxies accrete more metal-poor satellites. This flattening is not as prominent in the $M_*/R_e$ relations,
possibly because compact metal-rich satellites sink to the center of progenitors, whereas diffuse metal-poor satellites distribute around the outskirts \citep{boylan-kolchin_satellite_2007,amorisco_contributions_2017}. Thus, minor mergers may reinforce the $M_*/R_e$ relations \citep[see discussion in][]{barone_sami_2018}. 

\begin{figure*}[hbtp!]
    \centering
    \vspace{0.09in}
    \includegraphics[width=0.95\textwidth]{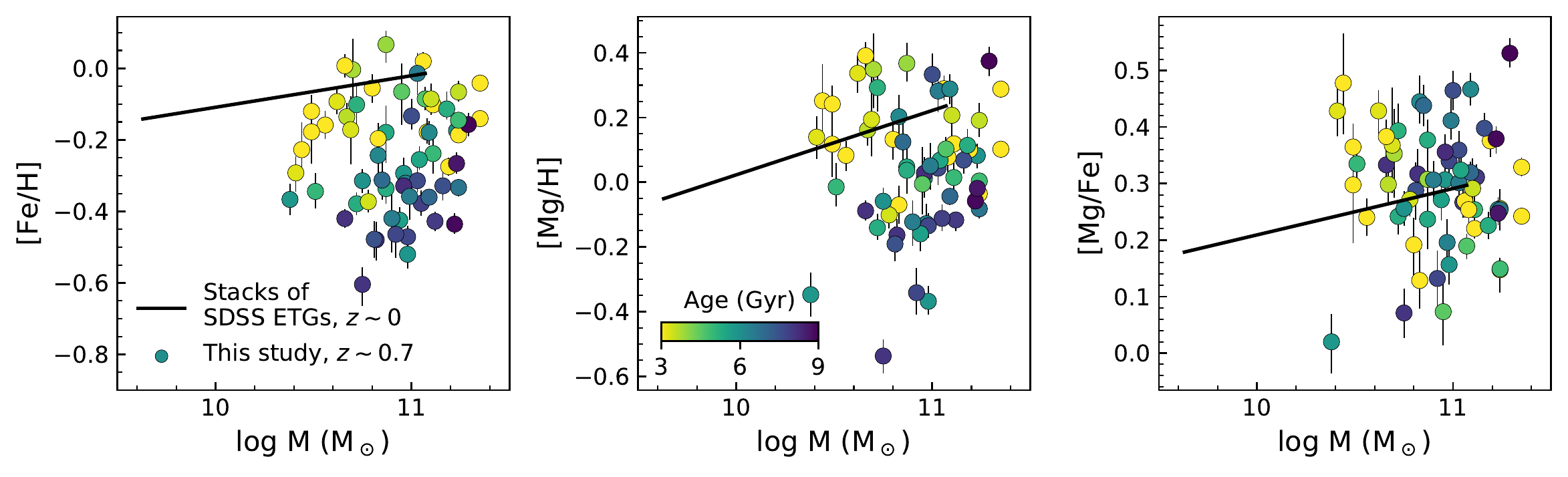}
    
    \caption{The mass-[Fe/H] (left), mass-[Mg/H] (center), and mass-[Mg/Fe] (right) relations colored by stellar age. The black lines show the results from stacks of local quiescent SDSS galaxies from \cite{conroy_early-type_2014}. We find $\sim0.2$ dex evolution in the mass-[Mg/H] and mass-[Fe/H] relations, and a clear trend between metallicity and stellar age at $z\sim0.7$, such that younger galaxies are more metal rich. We find no such trend or evolution in the mass-[Mg/Fe] relation.
    }
    \label{fig:compare}
\end{figure*}

Next, we look at [Mg/Fe] and stellar population age. Since [Mg/H] and [Fe/H] enrich on different timescales, [Mg/Fe] is a direct tracer of how quickly a galaxy formed its stars. [Mg/Fe] is often discussed in tandem with age, as the combination of the two quantities provides a strong constraint on when and how quickly galaxies formed their stellar mass. As shown in the bottom two rows of Fig.\;\ref{fig:ab}, [Mg/Fe] does not correlate with $M_*$ or $M_*/R_e$, while age marginally correlates with $M_*$. 

Our [Mg/Fe] and age results are similar to findings at low redshifts. At \zlo, the $M_*/R_e-$[Mg/Fe] and $M_*/R_e-$age trends are shallow and the mass-[Mg/Fe] and mass-age relations flatten above $\log M_*/M_\odot\gtrsim10.5$ \citep{mcdermid_atlas3d_2015,scott_sami_2017,barone_sami_2018}. Our results also agree with studies at $z\sim0.5$ that find little to no correlation between mass and [Mg/Fe], but a positive correlation between mass and age \citep{leethochawalit_evolution_2019,choi_assembly_2014,gallazzi_charting_2014}, and no correlation between galaxy age and size \citep[e.g.,][]{zanella_role_2016,fagioli_minor_2016}. Drawing from the same LEGA-C parent sample, \cite{wu_fast_2018} find that at a fixed mass, smaller quiescent galaxies actually tend to be older using D$_n$(4000), but they ignore the D$_n$(4000) metallicity dependence. In this study, we find a strong correlation between metallicity and galaxy size, thus explaining larger D$_n$(4000) for smaller galaxies. 

Our age results imply that the integrated SFHs of all the stars currently in a galaxy (both in situ and accreted) only slightly depend on mass, with more massive galaxies forming at higher redshift. Our [Mg/Fe] results indicate no clear trend between the star formation timescale and stellar mass. However, the interpretation is not straightforward, as both minor mergers and star formation timescales affect [Mg/Fe]. To break this degeneracy, studies need to push to higher redshifts, where the effect of mergers is less significant. Indeed, recent single-object studies have found that massive quiescent galaxies at $z\sim2$ are extremely $\alpha$-enhanced, a result that is in favor of evolution via minor mergers \citep{kriek_massive_2016,jafariyazani_resolved_2020}.

\begin{figure*}[!tp]
    \centering
    \includegraphics[width=\textwidth]{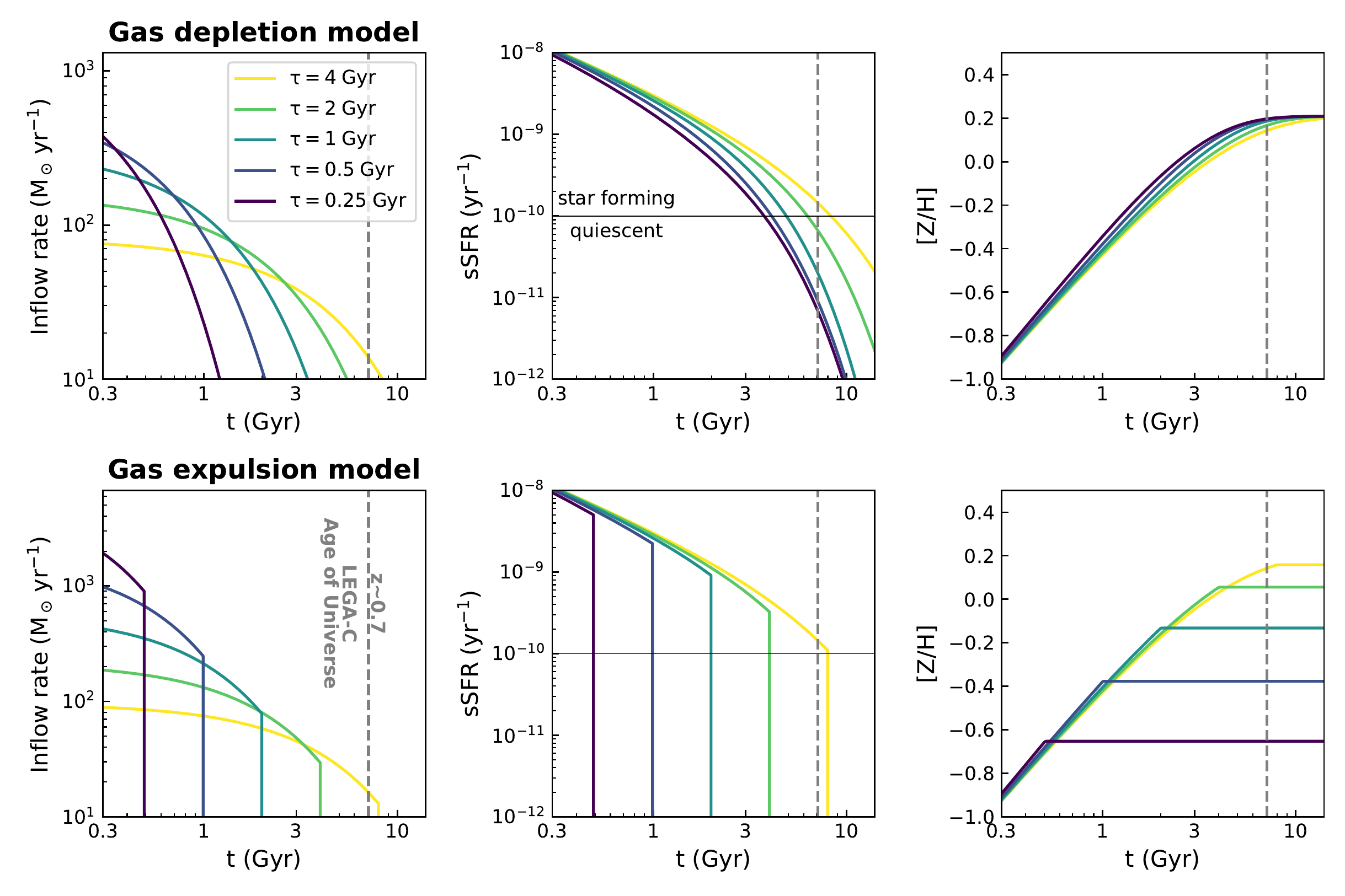}
    \caption{Results from a chemical evolution model with an exponentially declining inflow rate for five different inflow timescales, $\tau$. The left, middle, and right panels show the inflow rate, specific SFR, and mass-weighted stellar metallicity [Z/H], respectively. The top row is the ``gas depletion" model with solutions directly from \cite{spitoni_new_2017}. The bottom row is the ``gas expulsion" model where inflows from the \cite{spitoni_new_2017} model are terminated and the remaining star-forming material is removed after two $e$-folding times.
    Both models assume a star formation efficiency $\epsilon=0.5\,\rm Gyr^{-1}$, a \cite{chabrier_galactic_2003} IMF, and supernovae feedback with a mass-loading factor $\lambda=1$. The gas expulsion model successfully reproduces both the observed age-metallicity trend and the MZR evolution.
    }
    \label{fig:chemev_spitoni}
\end{figure*}

\section{Evolution in the Mass-Metallicity Relation}
\label{sec:evol}


Next, we investigate the evolution of the quiescent MZR between \zhi\ and \zlo. In Fig.\;\ref{fig:compare} we show [Mg/H], [Fe/H], and [Mg/Fe] as a function of stellar mass, colored by stellar population age. We compare these results with stacks of $z\sim0$ Sloan Digital Sky Survey (SDSS) galaxies from \cite{conroy_early-type_2014} that have been fit using the procedure in Section \ref{sec:alf}. We find that, at a given mass, there is $\sim0.2$ dex of evolution in both the mass-[Mg/H] and mass-[Fe/H] relations between \zlo\ and \zhi, and that younger galaxies have higher [Fe/H] and [Mg/H]. Interestingly, at $z\sim0.7$ \cite{gallazzi_charting_2014} find no evolution using Lick indices. However, the uncertainties on their metallicities are significantly larger than in this work. At $z\sim0.5$, \cite{choi_assembly_2014} and \cite{leethochawalit_evolution_2019} find that metal abundances are $\sim0.1$ dex lower using full spectral modeling, in broad agreement with our results.

Since quiescent galaxies are, by definition, no longer forming stars, they are unable to change their stellar metallicities by star formation. Instead, metallicities can change by either minor mergers or population growth (i.e., progenitor bias). The observed MZR evolution cannot be explain by minor mergers, as the accretion of low-mass metal-poor satellites would \textit{decrease} the metallicity of the progenitor galaxy. Progenitor bias, in contrast, can explain the observed evolution only if galaxies that quench at later times also have higher metallicities. Indeed in Fig.\;\ref{fig:compare} we observe that younger galaxies have higher [Mg/H] and [Fe/H]. 


To understand why galaxies that quench at later times have higher metallicities, we turn to chemical evolution models. We utilize solutions from \cite{spitoni_new_2017}, which assume an exponentially decreasing inflow rate with timescale $\tau$, a linear \cite{schmidt_rate_1959} law, and an outflow rate proportional to the star formation rate (SFR) with a mass-loading factor $\lambda$. In the first row of Fig.\;\ref{fig:chemev_spitoni} we show solutions for the inflow rate, SFR, and mass-weighted metallicity for model galaxies with five different inflow timescales and a constant mass-loading factor of $\lambda=1$. In the second row of Fig.\;\ref{fig:chemev_spitoni} we show the same five model galaxies, except the star formation and gas inflows are instantaneously terminated after two $e$-folding times (e.g., AGN feedback). Thus, the first model represents quenching via smooth gas depletion, while in the second model the quenching happens much more abruptly via gas expulsion\footnote{The star formation quenching does not need to be instantaneous, and can instead be modeled as a leaky box with a large mass-loading factor.}, in particular for early quenchers. In both models, the four galaxies with the shortest star-formation timescales would be identified as quiescent at the LEGA-C redshift, whereas the fifth and youngest model with the highest metallicity would only be added to the quiescent population at later times (thus identified as quiescent by $z=0$).

In the smooth gas depletion model, all quiescent galaxies end up with the same high metallicity by $z=0$ no matter their star formation timescale. Thus, this model fails to reproduce the observed evolution. One way to prevent the galaxies from converging to the same metallicity is to vary the mass-loading factor such that at a constant mass galaxies with the shortest inflow timescales have $10x$ more efficient outflows than galaxies with the longest inflow timescale. Given that the mass-loading factor is thought to be primarily driven by galaxy mass \citep[e.g.,][]{dekel_origin_1986}, and that here we are considering models at the same mass, this large difference in outflow efficiency is unlikely. 

In the gas expulsion model we find that the younger galaxies -- those that quench at later times -- are more metal enriched. Galaxies that quench at earlier times shut off their star formation when the inflow rate is still high and thus the gas-phase metallicity is still low. The galaxies that quench at later times, however, have a lower inflow rate and thus higher gas-phase metallicity when they quench, leading to a higher stellar metallicity. The [Z/H] panel shows that the gas expulsion model is successful at reproducing the observed age-metallicity trend found in Fig.\;\ref{fig:compare}. The same model can also explain the observed MZR evolution; galaxies that quench before \zhi\ have lower metallicities than those that quench at later times. Thus, in this model, the average metallicity of the quiescent population increases over cosmic time.






Finally, we turn to [Mg/Fe]. In the right panel of Fig.\;\ref{fig:compare} we find no [Mg/Fe] evolution between \zhi\ and \zlo, and no clear trend with age. This result is consistent with the work by \cite{leethochawalit_evolution_2019} and \cite{choi_assembly_2014} who find little to no [Mg/Fe] evolution out to $z\sim0.54$. The lack of [Mg/Fe] evolution is hard to interpret because it can be affected by both the star formation history and minor mergers. Both processes should decrease the mass-[Mg/Fe] relation over time; however, in reality, these effects are subtle due to the shallowness of the mass-[Mg/Fe] relation and the weakened sensitivity of [Mg/Fe] at longer star formation timescales. In order to disentangle both effects, we need to push current studies to higher redshifts, where the effects of mergers are less significant and the difference in star formation histories are easier to measure.

\section{Summary and Conclusion}

In this Letter we derive chemical abundances and stellar population ages for \num massive quiescent galaxies from the LEGA-C survey at $z\sim0.7$ and examine them as a function of mass and size. We find that [Mg/H] and [Fe/H] do not vary with $M_*$ but are correlated with $M_*/R_e$, reinforcing the picture that the strength of the gravitational potential was instrumental in regulating their chemical enrichment. We also find that age and $M_*$ are marginally correlated, but that [Mg/Fe] does not vary with $M_*$. These results suggest that star formation timescales are independent of mass and size but that more massive galaxies formed their stars earlier. 

Through comparison with stacks of $z\sim0$ quiescent galaxies from SDSS, we find that the mass-[Mg/H] and mass-[Fe/H] relations increased by $\sim0.2$\;dex since $z\sim0.7$. Furthermore, we find that older galaxies at $z\sim0.7$ have lower metallicities. The observed evolution and age trend can be explained in a chemical evolution model where galaxies quench via a major outflow event (e.g., AGN feedback). 

This study was enabled by the ultra-deep LEGA-C spectra, which, in combination with full-spectrum modeling techniques, has provided the most robust measurements of chemical abundances at $z\sim0.7$.
Our results highlight how ultra-deep spectra can be harnessed to uncover the star-formation and chemical-enrichment histories of quiescent galaxies, and further demonstrate that metallicities can be used to constrain star-formation quenching models \citep[see also][]{peng_strangulation_2015,spitoni_new_2017,trussler_both_2020}. In the future, ultra-deep ground-based surveys and JWST/NIRSpec will enable the extension of this analysis to higher redshift, allowing for better constraints on the quiescent MZR, its evolution, and its dependence on galaxy size.\\

We would like to thank Andrew Newman for useful conversations about \texttt{alf} fitting, Brittany Vanderhoof, Jeyhan Kartaltepe, and Caitlin Rose for sharing the COSMOS F814W PSF and useful discussion regarding size fitting, the LEGA-C team for making their dataset public, and the anonymous
referee for a constructive report. We acknowledge support from NSF AAG grants AST-1908748 and 1909942.

\software{\texttt{GALFIT} \citep{peng_detailed_2010}, FSPS \citep{conroy_propagation_2009}, \texttt{FAST} \citep{kriek_ultra-deep_2009}, \texttt{alf} \citep{conroy_stellar_2012,conroy_metal-rich_2018}}

\clearpage

\bibliography{zoterolib}
\end{document}